# Characterisation of the primary X-ray source of an XPS microprobe Quantum 2000

UWE SCHEITHAUER

*82008 Unterhaching, Germany*
Phone: + 49 89 636 44143
E-mail: scht.uhg@googlemail.com

## Abstract

The outstanding design feature of an XPS microprobe Quantum 2000 is the double focussing ellipsoidally shaped quartz monochromator of the X-ray source. This device monochromatizes the $Al_{k\alpha}$ radiation and refocuses the X-rays from the Al anode to the sample surface. This way, on one hand a variation of the diameter of the X-ray generating electron beam allows to vary the X-ray beam diameter on the sample surface. On the other hand a scanning of the electron beam across the Al anode scans the X-ray beam across the sample surface.

The X-ray source was characterized in detail. The lateral dependency of the primary X-ray intensity and the peaks FWHM were measured as function of the position within the electrostatically rasterable scan area. Additionally, the focussing quality of the monochromator was determined. Therefore the lateral intensity distribution within the primary X-ray beam was estimated far below the 1% intensity level.

## 1. Introduction

By modern XPS laboratory instruments a spatial resolution of ~ 10 μm is attainable at optimum. In commercial laboratory instrumentation this is achieved by two different design concepts: Either a fine-focused X-ray source with a double focussing quartz monochromator is used or the acceptance area of the energy analyser defines the lateral resolution [1, 2]. In both cases a deflection of electrons is used to get a 2-dimensional image of the sample surface. Either the exciting electron beam of the fine-focused X-ray source or the electrons, which are emitted from the sample, are deflected.

In this paper the primary X-ray source of an XPS microprobe Quantum 2000 was investigated. It has a fine-focused X-ray source with a double focussing ellipsoidally shaped quartz monochromator. The relative X-ray intensity and energy resolution of Cu2p3/2 peaks were measured as function of the lateral position of the X-ray beam within the electrostatically rasterable scan area on the sample surface. The performance of the double-focusing quartz monochromator in terms of the intensity distribution of the X-ray beam across its footprint at the sample's surface was investigated. Using apertures this quantitative lateral intensity distribution of the primary X-ray beam was measured far below the 1% intensity level.

## 2. Instrumentation

Fig. 1 shows a schematic drawing of the principal components of an X-ray microprobe Quantum 2000. The Quantum 2000 is an XPS instrument with a focused primary X-ray beam. The spatial resolution of an XPS microprobe Quantum 2000 is achieved by the combination of a fine-focused electron beam generating the X-rays on a water cooled Al anode and an ellipsoidally shaped mirror quartz monochromator, which monochromatizes and refocuses the X-rays to the sample surface [1, 3-5]. This way, the X-ray beam scans across the sample as the electron beam is scanned across the Al anode by electrostatic deflection. On the sample surface an area of ~ 1.4 * 1.4 mm$^2$ at maximum can be scanned by applying electrostatic deflection voltages to the electron beam. By controlling the electron beam diameter, nominal X-ray beam diameters of between ~5 μm and ~200 μm are selectable for the instrument used here [6]. X-ray induced secondary electron images localize sample features by using a rastered X-ray beam.

The monochromator adjustment was done and revised from time to time according to the manufacturer instructions. For a coarse adjustment, the X-ray intensity has to be maximized by a variation of the mechanical tilt-angle of the quartz monochromator. More often a fine-readjustment was achieved by a variation of monochromator heating temperature, which changes the lattice constant of the quartz crystal and thereby the Bragg angle. The adjustment aims to have the maximum intensity on the centre line of the electrostatically rasterable scan area with respect to the dispersive direction (y ≈ 0, see fig. 2).

Emittance matching: The analyser acceptance area is synchronized with the X-ray beam position on the sample by emittance matching. Voltages which are synchronized with the raster of the exciting electron beam are applied to electrostatic deflection plates at the analyser entrance for this purpose. Dynamic dispersion compensation: It compensates the energy variation of the primary monochromated X-ray beam while the beam position is shifted along the disperse to direction of the monochromator. This is due to slight deviations from the optimum Bragg angle while the beam position is shifted along the disperse direction. To compensate this energy variation, the retarding potential of the analyser input lens varies according the disperse direction raster position of the X-ray beam.

To enhance the instruments sensitivity, the energy analyser is equipped with a multichannel electron detector. It uses 16 discrete channels for parallel detection. The geometrical lateral distance of these separate detector channels defines the possible combination of the energy analyser pass energy settings and binding energy grid point intervals. Ten electron analyser pass energies between 2.95 eV and 187.85 eV are selectable [6].

For all samples measured here in the Quantum 2000 the incoming X-rays are parallel to the surface normal. In this geometrical situation, the average geometrical energy analyser take-off axis and the differentially pumped $Ar^+$ ion gun are oriented ~ 45° relative to the sample surface normal. The ion gun is used for charge neutralization [7], sputter cleaning and depth profiling of the samples.

The instrument is operated in an air-conditioned temperature stabilized laboratory. To avoid thermal drifts of the electronics all components are continuously switched on. Electrical components are only powered off for service issues.

All measurements presented here were made using the original instruments configuration, which was not altered to a great extent since its installation in the year 2001. Only the quantitative lateral resolution measurements with the new ellipsoidally shaped mirror quartz monochromator were recorded after an upgrade in the year 2010, obviously.

## 3. Samples

A Cu pure metal foil and Pt apertures usually used in electron microscopes, were utilized for the measurements presented here. Before the measurements the samples were cleaned by $Ar^+$ ion sputtering.

## 4. Experimental Characterization of the X-ray source

The Quantum 2000 uses a primary X-ray beam with a variable beam diameter and the X-ray beam is rasterable on the sample. Hence the relative X-ray intensity and the energy resolution / peak shape were measured as function of the lateral position of the X-ray beam on the sample within the area, which is covered by electrostatic deflection of the X-ray generating electron beam. The performance of the double-focusing mirror quartz monochromator was investigated by measuring the lateral intensity distribution within the primary X-ray beam [8].

## 4.1 Relative X-ray Intensity as Function of Lateral Position

The relative X-ray intensity was measured using the photoemission intensity of a sputter cleaned Cu sample. Fig. 2 shows a mapping of the Cu 2p3/2 signal intensity over an area of ~ 1.3 * 1.3 mm$^2$, which was scanned by electrostatic rastering of the X-ray generating electron beam. The x-axis is parallel to the non-dispersive and the y-axis is parallel to the dispersive direction of the monochromator, respectively. The highest intensity is observed for negative x-values. In this non-dispersive direction a deviation of +5% and -10% from the mean value was determined. A slight misalignment of the Bragg angle is detectable. The intensity is highest for small negative y-values. Since the Bragg angle fine alignment is done by the monochromator temperature via a lattice parameter variation of the quartz crystal, a temperature modification by a few degrees would correct this and relocate the intensity maximum to y = 0. As expected, the intensity decreases for non-optimum Bragg angels in the dispersive y-direction. A decrease by a factor of ~ 7 for beam positions with a maximum deviation from the optimum Bragg angle was measured.

The Cu 2p3/2 signal intensity shown in fig. 2 is proportional to the primary X-ray beam intensity. Additionally it is a convolution with the energy analyser transmission, which may be a function of the lateral position within the electrostatically rasterable scan area on the sample surface. This analyser transmission function was investigated by the measurements shown in fig. 3. The graph depicts the intensity ratio of the Cu3p signal divided by the Cu2p3/2 signal as function of the lateral position of the primary X-ray beam. Deviations by +12% and –7% from the mean intensity ratio are observed. The ratio has the highest values near the x-axis for negative x-values. It decreases for positive x-values and even more for higher absolute y-values. Due to the division of the two peak intensities the ratio is independent from the lateral intensity distribution of the primary X-ray beam. Therefore the plot shows the lateral dependency of the analyser transmission function. But the two Cu peaks used here have a binding energy difference of ~ 854 eV. So we have not only the energy analyser transmission as function of the lateral position within the electrostatically rasterable scan area but also a convolution with the energy dependence of the energy analyser transmission, which may be a function of the position within the electrostatically rasterable scan area. This way the results of fig. 2 overestimate the contribution of the analyser transmission function regarding the dependency form the position within the electrostatically rasterable scan area.

In summary the results presented in fig. 2 and fig. 3 demonstrate, that the contribution of the analyser transmission to the intensity variation is negligible compared to the drastically intensity decrease by a factor 7 for non optimum Bragg angles.

## 4.2 Energy Resolution and Peak Shape as Function of Lateral Position

Fig. 4 shows the energy resolution as function of the lateral X-ray beam position. The energy resolution was determined using a measurement of the FWHM (full width at half maximum) of the Cu 3p3/2 signal. The data were measured with a low energy analyser pass energy giving high energy resolution of the signals. The FWHM estimated by peak fitting is coded by the circle diameter on a linear scale. The colours of the circle area gives the relative peak intensity as discussed above. For the (0 μm / 0 μm) and (±600 μm / ±600 μm) positions the measured spectra are inserted. The FWHM increases significantly for peaks measured at negative y-positions. Asymmetric signals, which have to be fitted by two peaks, are observed for the position (-600 μm / -600 μm) and (600 μm / 600 μm).

### 4.3 Quantitative Lateral Resolution

The knowledge of the quantitative lateral resolution is essential, if the XPS instrument is used to analyse compositions of small features such as Al bond pads of microelectronic devices, for instance. Bond pads of microelectronic devices are in the order of 70 * 70 μm$^2$. Regarding unwanted contamination on the bond pad surface, signal contributions from outside the bond pad due to long tail intensity distributions of the primary X-ray beam have to be taken into account. The knowledge of the instruments quantitative lateral resolution function allows judging whether a detected trace element is present on the Al surface or if it might be due to a signal contribution from outside the bond pad. More generally, due to these signals belonging to long tails of the primary X-ray beam intensity distribution, the detection limit of trace elements on small features depends on the feature size if the same elements are present in the surrounding of the feature.

Improving the approaches reported in literature [9, 10], circular Pt apertures used in electron microscopes with different diameters were utilized as test samples for this analysis [8]. The insert in fig. 5 shows a schematic sketch of the apertures mounting. For different reasons using apertures as test samples is an excellent choice. No assumption about the primary X-ray beam profile is necessary, since a deconvolution of the measured signal is not required. For instance, this would have to be done for test samples, which uses distinct edges between two different materials, if the X-ray beam is scanned over this edge for its beam profile analysis. Independent from the exact in-plane X-ray beam profile the long tail contributions to the signal in all directions are measured simultaneously when the primary X-ray beam is positioned in the centre of an aperture. Obviously the hole of an aperture is contamination-free. A material redeposition by a manufacturing process or by the Ar$^+$ sputter cleaning to a hole is impossible [11].

The measurement results are depicted in fig. 5. Against the aperture diameter the normalised Pt4f signal measured in the aperture centre is plotted on a logarithmic scale. For the signal normalisation a second measurement on Pt in an adequate distance to the hole is utilized. All measurements were done near to the centre of the electrostatically rasterable scan area. Please notice that the y-axis shows only the signals up to 10% of the maximum intensity. The data were measured for 4 primary X-ray beam diameters using both, the old and, after a system upgrade, the new monochromator crystal. The X-ray beam sizes are given by the insert of fig. 5. Following the manufactures approaches [5], they were estimated using the 80% to 20% signal variation when scanning the primary X-ray beam over a material edge. Fig. 5 shows, that for larger apertures the normalized Pt signals are independent from the primary X-ray beam diameter. The normalized intensity values of the old monochromator crystal are significantly higher than the values of the new monochromator crystal. The quantitative lateral resolution is defined by the aperture diameter of that aperture, for which by a measurement in the centre of the aperture hole the Pt signal intensity is 1% of the reference intensity measured on solid Pt. Using this definition, the quantitative lateral resolution is ~450 μm for the old monochromator. After the monochromator upgrade the quantitative lateral resolution improves to ~190 μm.

Going back to the 70 * 70 μm$^2$ bond pad measurement mentioned above: Even with the new monochromator ~ 3 % of the primary X-ray intensity will impinge on the bond pads surroundings and initiate photoelectron emission from this area. These 3% have to be taken into account when interpreting low intensity XPS signals attributed to unexpected elements. Whether these contaminations are on the bond pad surface or are contributions of the bond pad surroundings is indeterminable.

## 5 Conclusions

The double focussing ellipsoidally shaped quartz monochromator is the outstanding design feature of the Quantum 2000. Using focusing and electrostatic deflection of the X-ray generating electron beam, it enables variable X-ray beam sizes and a scanning of the X-ray beam on the sample surface over an area of 1.4 * 1.4 mm$^2$ laterally. Measurements of the lateral dependency of the relative primary X-ray intensity, the XPS peaks FWHM / peak shape and the analyser transfer function were performed. Additionally, the quantitative lateral intensity distribution within the primary X-ray beam far below the 1% intensity level was estimated. This way imported features of the primary X-ray source were characterized.

The relative primary X-ray intensity varies in non-dispersive direction of the monochromator by ~ 20%. As expected, in the dispersive direction the intensity decreases by a factor of ~ 7. If the XPS intensity data are used to compare different specimen, for instance, it is crucial to measure all spectra on the same position within the 1.4 * 1.4 mm$^2$ electrostatically rasterable scan area to avoid misleading results due to primary X-ray intensity variations.

The lowest XPS peak FWHM of high energy resolution peak measurements is obtained for the optimum Bragg angle. In the non-dispersive direction no changes of the FWHM are observed for this angle. The FWHM nearly doubles for positions in negative dispersive direction. Asymmetric peaks were observed at two corners of the electrostatically rasterable scan area. Therefore measurements with high energy resolution of XPS peaks for chemical peak-shift analysis should be recorded at the optimum Bragg angle.

For measurements of small features the detection limit of trace elements is not determined by the FWHM of the primary X-ray beam but by the long tail intensity distributions of the primary X-ray beam. The quantitative lateral resolution of the instruments, which is defined by the 1% signal contribution level, depends on the monochromator quality. After a monochromator upgrade the quantitative lateral resolution improves from ~450 μm for the old monochromator to ~190 μm for the new one.

## Acknowledgement

All measurements were done using the Quantum 2000, instrument no. 78, installed at Siemens AG, Munich, Germany. I acknowledge the permission of the Siemens AG to use the measurement results here. For many fruitful discussions I would like to express my thanks to my colleagues and to the people of Physical Electronics, in particular to Mr. U. Roll, Mr. M. Schleich and Mr. T. Groß.

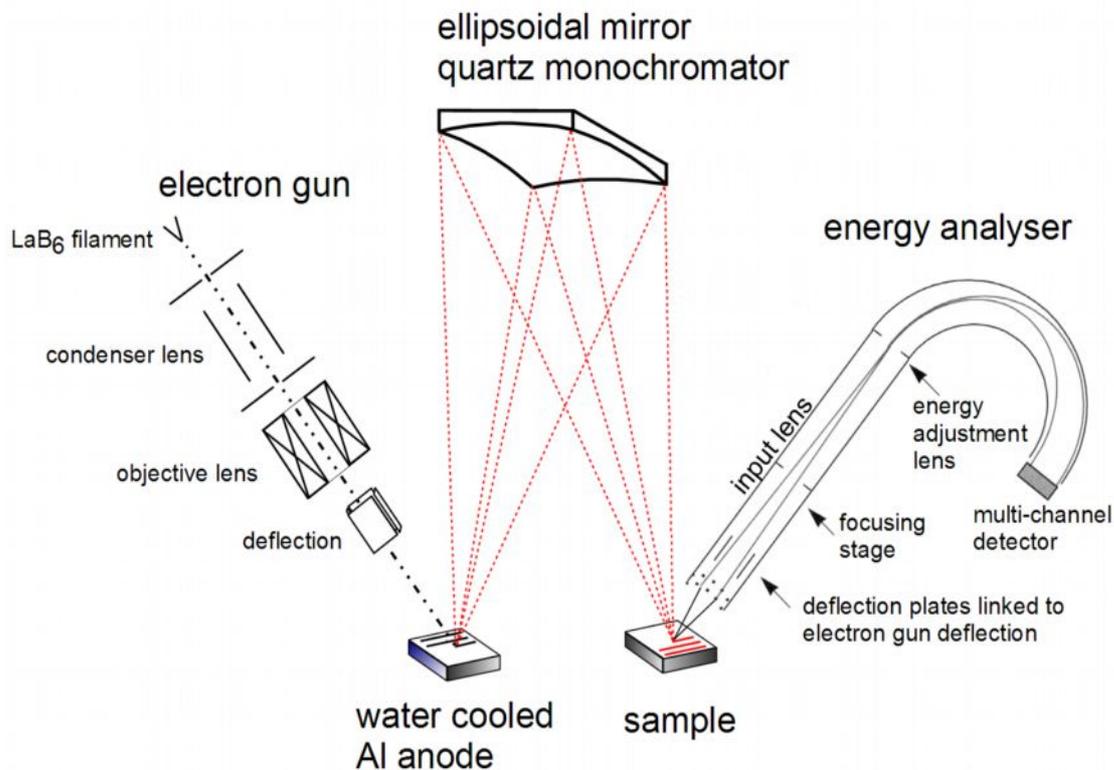

Fig. 1: schematic drawing of the principal components of a Quantum 2000 X-ray micro probe

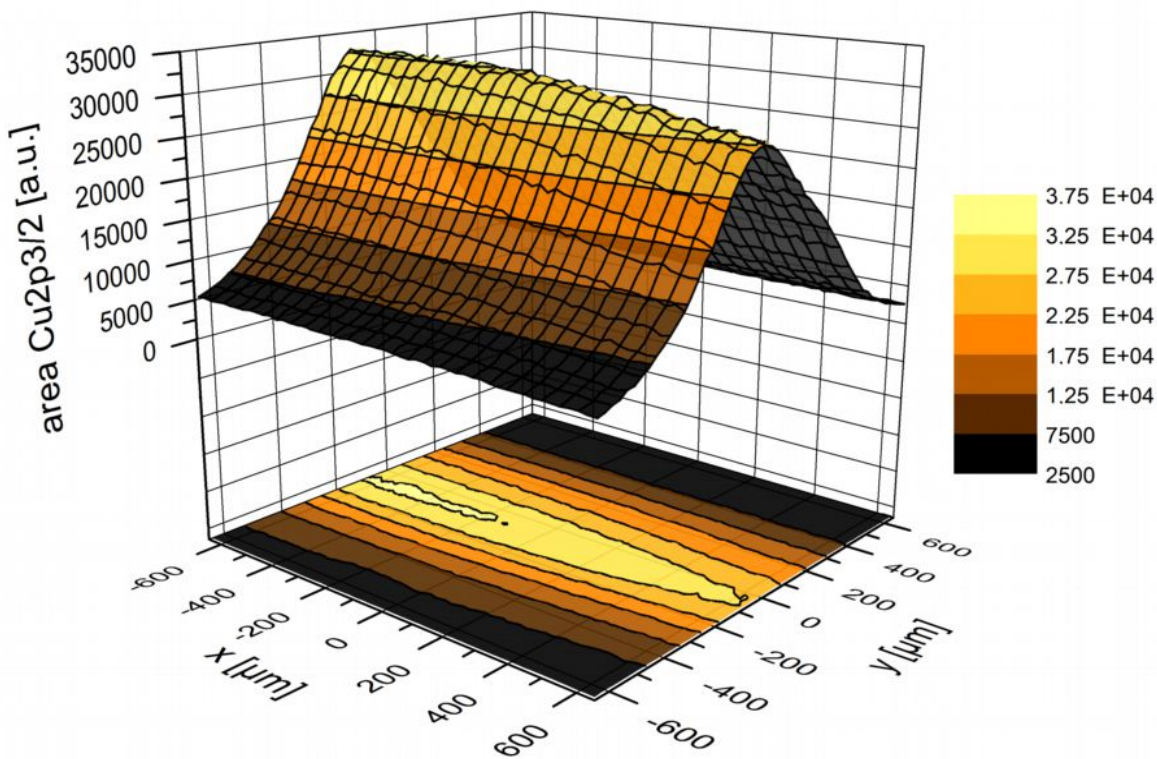

Fig. 2: intensity of the Cu2p3/2 peak as function of lateral position of the primary X-ray beam, X-ray beam diameter: 100 μm, pass energy: 93.9 eV

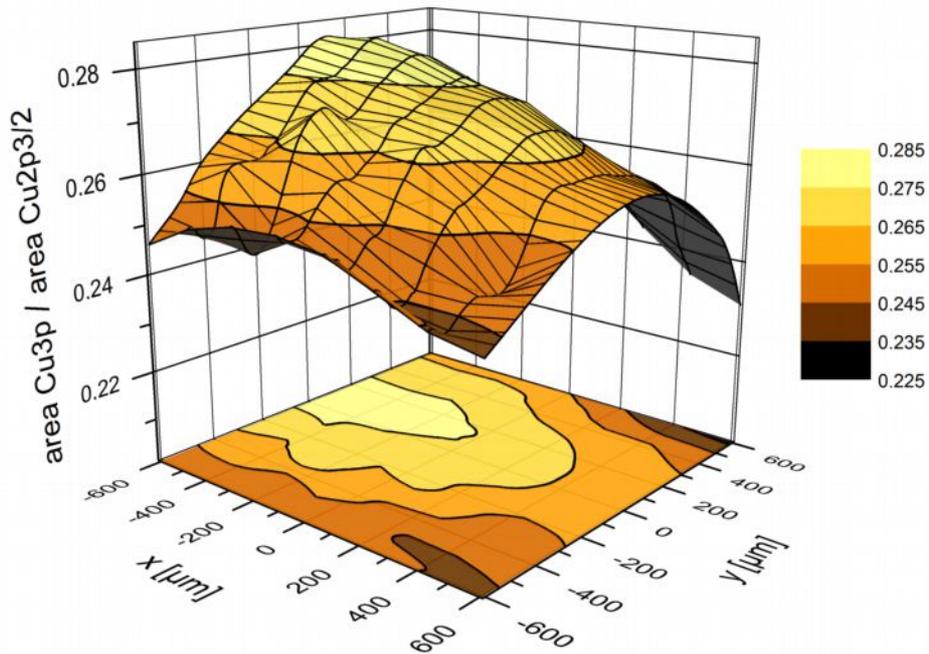

Fig. 3: intensity ratio of Cu3p peak to Cu2p3/2 peak as function of lateral position of the primary X-ray beam, X-ray beam diameter: 50 µm, pass energy: 93.9 eV

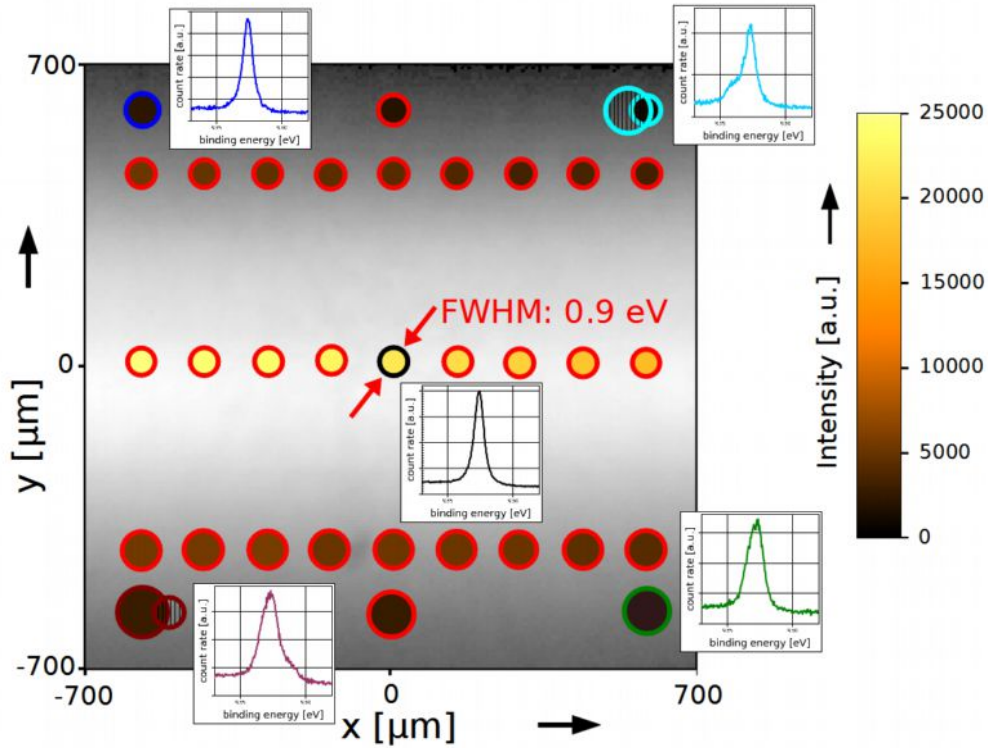

Fig. 4: FWHM of the Cu2p3/2 peak as function of lateral position of the primary X-ray beam, X-ray beam diameter: 50 µm, pass energy: 11.75 eV
The circle diameter gives the FWHM on a linear scale and the circle colour codes the intensity of the Cu2p3/2 signal. The inserted graphs show the measured Cu2p3/2 signal at (0µm / 0µm) and (±600µm / ±600µm).
The background shows an X-ray induced secondary electron image of the Cu sample.

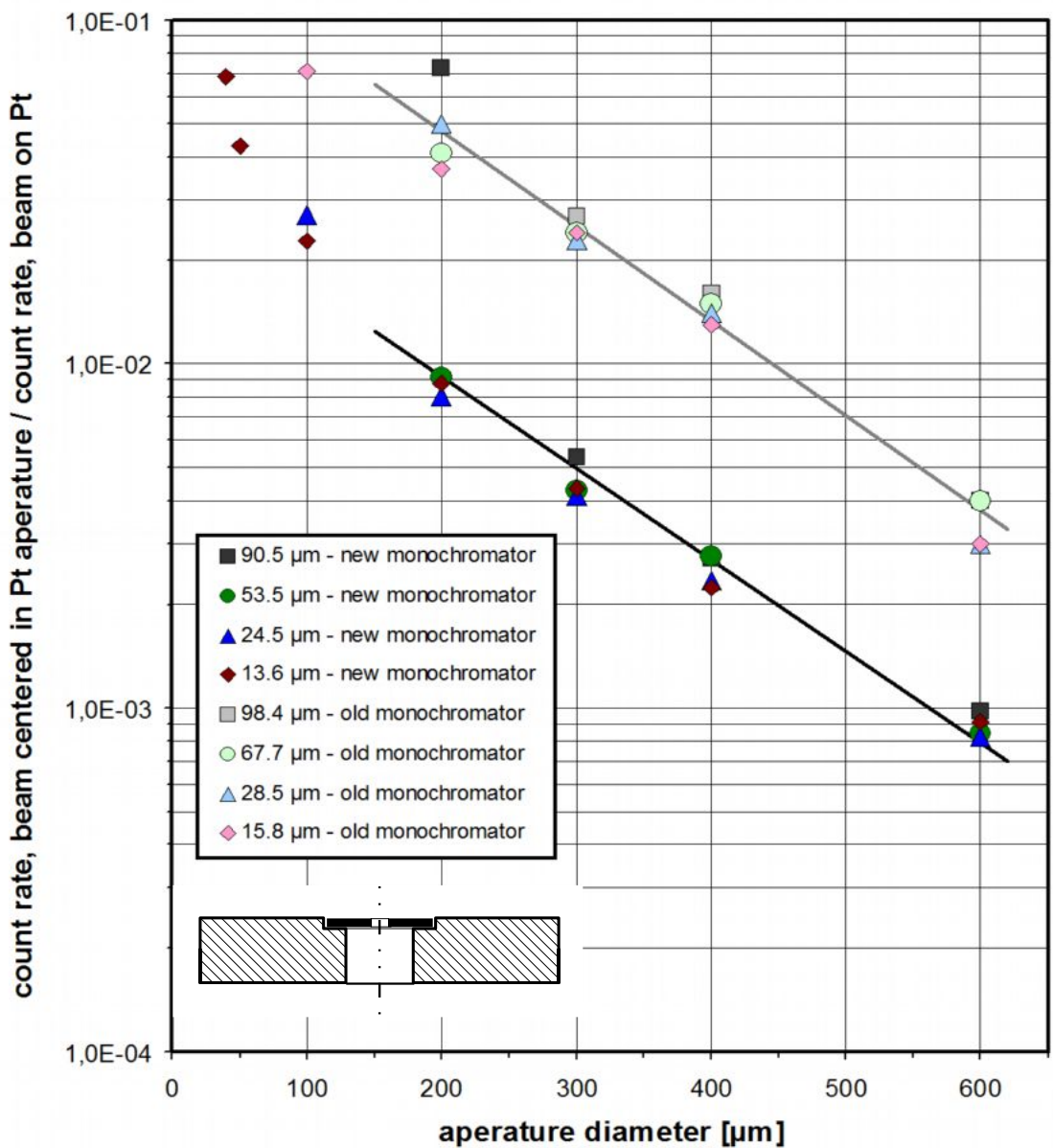

Fig. 5: count rate ratio as function of aperture diameter for 4 different primary X-ray beams, data measured with the old and new monochromator